\documentclass[12pt]{article}
\usepackage{latexsym}
\newcommand{\be}{\begin{equation}}
\newcommand{\ee}{\end{equation}}
\newcommand{\ba}{\begin{eqnarray}}
\newcommand{\ea}{\end{eqnarray}}
\newcommand{\sectn}[1]{\section{#1}\setcounter{equation}{0}}
\renewcommand{\thesection}{\arabic{section}.}
\renewcommand{\theequation}{\thesection\arabic{equation}}

\newcommand{\rca}{\mathcal{R}}
\newcommand{\Agrav}[1][]{\ensuremath{\mathcal{A}_{grav}^{#1}}}
\newcommand{\dl}[1]{\ensuremath{\partial}_{#1}}
\newcommand{\cov}[1]{\ensuremath{\nabla}_{#1}}
\newcommand{\half}{\ensuremath{\frac{1}{2}}}
\newcommand{\met}[2]{\ensuremath{g_{#1#2}}}

\newcommand{\ein}[3][]{\ensuremath{\mathcal{E}_{#2#3}^{#1}}}
\newcounter{supp}
\newcommand{\bleteq}{\setcounter{supp}{\value{equation}}
\stepcounter{supp}
\setcounter{equation}{0}
\renewcommand{\theequation}{\thesection\arabic{supp}\alph{equation}}}
\newcommand{\eleteq}{\setcounter{equation}{\value{supp}}
\renewcommand{\theequation}{\thesection\arabic{equation}}}

\begin{document}

 \rightline{EFI 97-42}

\thispagestyle{empty}
 \begin{center} \large {\bf Gravitational Analogues of 
Non-linear Born Electrodynamics}

 \medskip\normalsize 
James A. Feigenbaum, Peter G.O. Freund and Mircea Pigli\\

             {\em Enrico Fermi Institute and Department of Physics\\
             The University of Chicago, Chicago, IL 60637, USA}

 \end{center}
 \bigskip
 \bigskip
\bigskip \bigskip \bigskip

\noindent{\bf Abstract}: Gravitational analogues of the nonlinear 
electrodynamics of Born and of Born and Infeld are introduced and 
applied to the black hole problem. This work is mainly devoted to the 
2-dimensional case in which the relevant lagrangians are nonpolynomial in the 
scalar curvature.

\newpage
\thispagestyle{empty}
\mbox{}
\setcounter{page}{0}
\newpage
\bigskip
\sectn{Introduction}
\bigskip

In open string theory the dynamics of gauge fields 
on D-branes is governed \cite{P, L} by a Born-Infeld \cite{BI} lagrangian
or a non-abelian generalization thereof, with the corresponding point-charge 
solution
describing the way fundamental strings attach to branes \cite{CM, G}.
The Born-Infeld lagrangian is one in a class
devised by Born \cite{B} to remove the point charge singularity which mars 
classical electrodynamics. There is plenty of 
evidence that string theory leads to a finite quantum theory of gravity.
It has long been expected that quantum effects will ultimately remove the 
singularities of classical gravity and cut off curvatures at the string scale.
Were one to integrate out all other degrees of 
freedom, an effective lagrangian nonpolynomial in {\em curvature} components
would arise. One may therefore wish to investigate nonpolynomial 
lagrangians with the feature that they
cut off the gravitational field, i.e. the space-time Riemann
curvature components, in a manner similar to that in which the nonpolynomial
Born and Born-Infeld lagrangians cut off the electric
field components, i.e. the curvatures of the principal fibre bundle 
corresponding to electrodynamics or to one of its non-abelian generalizations.

Born achieves this through lagrangians which have a branch point in some
curvature scalar (e.g $F_{\mu\nu} F^{\mu\nu}$), so that only for values of this
invariant smaller than a critical value (determined by the position of the 
branch point) is the lagrangian real and thus meaningful. The simplest 
examples of a Born lagrangian are
\begin{equation}
{\cal L}=B(F_{\mu\nu} F^{\mu\nu}), 
\ee
with
\be
B(x)= \Lambda^2\Bigl[\sqrt{1-\frac{x}{2\Lambda^2}} -1\Bigr],
\ee
or 
\be
B(x)=\frac{\Lambda^2}{2}\ln({1-\frac{x}{2\Lambda^2}}).
\ee
In both these examples the branch point is located at
$F_{\mu\nu} F^{\mu\nu}=2\Lambda^2$, so that the square of the electric field
$E^2$ is cut off, $E^2\leq \Lambda^2$, and cannot then blow up
at the location of the point charge.
 
Another example is the Born-Infeld lagrangian
\be
{\cal L}_{BI}= -\Lambda^2 \sqrt{-\det(g_{\mu \nu} + \frac{F_{\mu \nu}}{\Lambda})}.
\ee
Suitable traces in internal space are to be performed in the non-abelian case
\cite{TS}.

Similar constructs with scalars built out of Riemann tensor components 
instead of electromagnetic field components will be considered in Sections 2
and 3.
For simplicity we will treat in this paper the 2-dimensional case. 
At the end of Section 2 we  
show how this can  be generalized to higher dimensions.
In Sections 4-6 we take up the black hole problem in the 2-dimensional case.
Unlike the $R$-linear case, this is no longer analytically soluble (with or 
without a dilaton field), but we can still derive its main physical features.
We will see that the black hole singularity is eliminated and inside the event
horizon a geodesically complete, asymptotically de Sitter space emerges. 
This occurs because the
appearance of space-time singularities associated with black holes is not a
consequence of the principle of general covariance which underlies general 
relativity, but rather of the specific $R$-linear Einstein-Hilbert lagrangian.
In Section 7 we will discuss our results.

Our work is related to the interesting papers of Brandenberger \cite{B1, B2}
who uses Lagrange multipliers to eliminate curvature singularities. 
A Born-Infeld gravity of a very different kind was  considered in \cite{PAL}.

\bigskip
\bigskip

\sectn{$R$-nonlinear Gravitational Lagrangians}
\bigskip

By switching to a lagrangian nonpolynomial in the electromagnetic field, 
Born was able to eliminate
the singularity of the electric field at the position of a point charge. 
In a similar vein, we shall construct lagrangians nonpolynomial in the 
components of the Riemann curvature, which eliminate the 
curvature singularity associated with a black hole solution. To keep things 
simple we will mainly study the 2-dimensional case where the scalar curvature
completely describes the situation. There are black holes in this case as 
pointed out by Witten \cite{W} for 2-d dilaton gravity and also studied in the
important work of CGHS \cite{CGHS} and of Wadia et al \cite{WA}. We will first 
construct the nonpolynomial counterparts of 2-d dilaton gravity. The original
``polynomial" version of this theory is governed by the action \cite{CGHS}
\be
{\cal A}_{DG}= \int d^2 x \sqrt{-g} e^{-2\phi}[4\lambda^2 + R + 4(\nabla \phi)^2],
\label{cghs}
\ee
where $\phi$ is the dilaton field and $\lambda$ the cosmological constant.
The field equations obtained by varying this action have the black hole 
solution
\bleteq
\be
g_{uu}=g_{vv}=0, ~~~ g_{uv}=g_{vu}= -\frac{1}{2} e^{2\rho} 
\label{me}
\ee
with
\be 
e^{2\rho}=\frac{1}{\frac{M}{\lambda} - \lambda^2 uv}
\label{WW}
\ee
and
\be
\phi(u,v)=\rho(u,v). 
\label{WW1}
\ee
\eleteq
The curvature scalar
\be
R=8 e^{-2\rho}\partial_u \partial_v \rho
\label{R}
\ee
obtained from the black hole metric (\ref{WW}) is then
\be
R=\frac{\frac{4M}{\lambda}}{\frac{M}{\lambda^3} - w}, 
\ee
where 
\be
w=uv,
\ee
so that $R$ is singular at 
\be
w=\frac{M}{\lambda^3},
\ee
the familiar 2-d black hole singularity \cite{W,CGHS,WA}. 
To maintain the role of the field $\phi$ as dilaton field and
in particular the linear dilaton flat-space solution, the only change we shall make will
be to replace, \`{a} la Born, the curvature term $\sqrt{-g} e^{-2\phi} R$
in the action Eq.~(\ref{cghs}) 
by $\sqrt{-g} e^{-2\phi} F(R)$, where $F(R)$ is a function 
with branch points at values $\pm \frac{8}{\beta}$ 
of the scalar curvature, so that
for $|R|\geq |\frac{8}{\beta}|$ the action is no longer real (here the factor
8 is inserted for future convenience).
The condition that the 
classical action be real thus cuts the scalar curvature off and no singularity
can arise. 

Specifically we will choose
\be
F(R) = 4\Biggl(\gamma+\Bigl(\frac{2}{\beta}-\frac{\gamma}{2}\Bigr)\sqrt{1+\frac{\beta}{8}R}-\Bigl(\frac{2}{\beta}+\frac{\gamma}{2}\Bigr)\sqrt{1-\frac{\beta}{8}R}\Biggr)
\label{fni}
\ee
where the coefficients  have been so chosen that
for $\beta \rightarrow 0$ 
the function $F(R)$ should reduce to the CGHS form $R$.
Instead 
of the square root branch point, we could equally well have chosen a logarithmic branch point.
For later use, let us write down the nonpolynomial 
action 
\be
{\cal A}_{F}= \int d^2 x \sqrt{-g} e^{-2\phi}[4\lambda^2 + F(R) + 4(\nabla \phi)^2],
\label{np} 
\ee
with $F(R)$ as in Eq.~(\ref{fni}).

In Section 6 we shall discuss the black hole problem based on this action.
Though string theory suggests a dilatonic gravity action, it may be of some 
interest to consider
in 2-d a purely gravitational action with black hole solution  (obviously this
cannot be the Einstein-Hilbert action which in 2-d is topological). This action
is most readily found by considering a general action of the form
\be
{\cal A}_{grav}^{(0)}= \int d^2 x \sqrt{-g} K(R)
\ee
with the function $K(R)$ to be determined by
writing down the corresponding field  equations and requiring that the black
hole metric Eq.~(\ref{WW}) be a solution of these equations. 
The field equations
then become differential equations for the function $K(R)$ which appears in 
the lagrangian. As shown in the Appendix, solving this differential equation 
yields the action
\be 
{\cal A}_{grav}^{(0)}= \int d^2 x \sqrt{-g} R \ln R.
\label{ln}
\ee
Actually two extra terms are possible: the classically irrelevant 
topological Einstein-Hilbert term and 
a term which involves a transcendental function of $R$, which however 
does not
share the ``residual Weyl invariance" of the action Eq.~(\ref{ln}) which will 
be described in Section 3. For this reason and for reasons of simplicity this 
term can be eliminated.
The action Eq.~(\ref{ln})
has some interesting properties which will be studied in the next 
section. Here we limit ourselves to presenting for this action a modification 
of the same type as that undertaken above for dilaton gravity. 
The modified action we propose is 
\be
{\cal A}_{grav}^{(\beta)} = \int d^{2}x \sqrt{-g} R [\ln R + 
        \beta \ln (a-R)]. 
\label{mp}
\ee
Notice that the modification here is obtained by adding a new term to 
the original action ${\cal A}_{grav}^{(0)}$ of Eq.~(\ref{ln}). In fact here 
${\cal A}_{grav}^{(0)}$ 
itself is already nonpolynomial in $R$, and had 
we also modified the original
$R\ln R$ term to say $R\ln (R+b)$ we would have eliminated the possibility of 
asymptotically flat solutions, as
will be shown in Section 4. 
The action (\ref{mp}) will be studied in detail in Sections 4 and 5.
In the next section we first describe the features of the action   
${\cal A}_{grav}^{(0)}$.

For the 2-d action (2.8) the scalar curvature is cut off at
$|R|\leq |\frac{8}{\beta}|$, whereas for the 2-d action (2.11) the cut-off is
$0\leq R\leq |a|$. In both these cases and in the 
more realistic higher dimensional case, the cut-off in the 
relevant curvature invariants is expected to occur near the string  scale.

To implement a similar program for higher dimensions, we have to contend 
with the facts that

a)  in addition to the scalar curvature and its powers, further scalar invariants can be formed from the Riemann tensor, 
e.g. $R_{\mu\nu\rho\sigma}R^{\mu\nu\rho\sigma}$, and

b)  the Einstein-Hilbert action ${\cal A}_{EH}$ is not topological.

An action which reduces to ${\cal A}_{EH}$ as the parameter 
$\beta \rightarrow 0$
and shares the features discussed above is
\be
\mathcal{A} = \int d^dx \sqrt{-g}\Bigl[~R + \beta \sqrt{1-\eta_1 R_{\mu\nu\rho\sigma}R^{\mu\nu\rho\sigma} - \eta_2 R_{\mu\nu}R^{\mu\nu} - \eta_3 R^2}~\Bigr]
\ee
with $\beta$ and $\eta_i~ (i=1,2,3)$ suitable parameters.
The dilatonic case can also be treated along these lines. 
In higher dimensional cases actions that only depend on the scalar curvature 
are of no use, as it is not the scalar curvature but 
$R_{\mu\nu\rho\sigma}R^{\mu\nu\rho\sigma}$ 
which becomes singular for the black hole.

\bigskip

\sectn{The $R\ln R$ lagrangian}
\medskip

As shown in the Appendix, the action Eq.~(\ref{ln}) is obtained 
by setting the constants $C_1$ and $C_3$ to zero in
the most general 2-d dilatonless 
gravity action which admits the black hole solution  
Eqs. (\ref{me}), (\ref{WW}).
The $C_3$ term, being 
topological, is classically irrelevant, as already noted above. Here we point
out a residual Weyl invariance which is realized if we also require $C_1~=~0$.

To this end notice that the action (\ref{ln}) is invariant under those Weyl 
transformations $g_{\mu \nu} \rightarrow exp(2\alpha) g_{\mu \nu}$ for which 
the function $\alpha$ is a solution of 
the 2-d wave equation $\Box \alpha=0$. To see 
this, write the metric in the standard form (\ref{me}).
Then a Weyl transformation amounts to a shift $\rho~\rightarrow~\rho~+~\alpha$.
Under this shift the change in the lagrangian density $\sqrt{-g} R \ln R$ 
is $-2\alpha \partial_{u} \partial_{v}\rho~ $ plus 
terms which vanish if $\alpha$
obeys the 2-d wave equation 
$\Box \alpha= -4 e^{-2\rho}\partial_{u} \partial_{v}\alpha = 0$. By two partial
integrations, the new term can be recast as the sum of a surface term and of a 
term proportional to $\partial_{u} \partial_{v}\alpha$ which again vanishes. So
we have established that 
the action (\ref{ln}) is invariant under these residual Weyl transformations 
\be
g_{\mu \nu} \rightarrow exp(2\alpha) g_{\mu \nu}, ~~~
\Box \alpha = 0.
\label{weyl}
\ee
This special Weyl invariance is lost when we  include the $C_1$ term in the 
action (\ref{eq:ares}). Requiring that this residual Weyl invariance be observed can 
thus be used to eliminate this $C_1$ term from the general action (\ref{eq:ares}), which
then yields uniquely the action (\ref{ln}).

The field equations obtained by extremizing 
the action (\ref{ln}) are
\begin{equation}
\ein[{(0)}]{\mu}{\nu} \equiv \half \met{\mu}{\nu} R 
	- \cov{\mu}\cov{\nu} \ln R
	+ \met{\mu}{\nu} \nabla^{\lambda}\cov{\lambda} \ln R = 0.
\label{eq:RlnRmo}
\end{equation}
For a metric of the type Eq.~(\ref{me}), we first concentrate on
the trace of the field equations, $\ein[{(0)}]{u}{v}=0$, which takes the form
\be 
\partial_{u} \partial_{v}(\ln \partial_{u} \partial_{v} \rho - 4 \rho)=0
\label{mircea}
\ee
It is worthwhile to notice that, as a consequence of this equation, 
the logarithm of the scalar curvature obeys the Liouville equation:
\be 
e^{\ln R} + \nabla _c \nabla ^c \ln R = 0.
\ee

Without any further ado the fourth order equation \ref{mircea} can be 
integrated twice  and one obtains the second order equation
\be
\partial_{u} \partial_{v} \rho= U_1(u) V_1(v) e^{4\rho},
\label{orig}
\ee
where $U_1(u)$ and $V_1(v)$ are two arbitrary functions. Defining the function
\be
\psi(u,v)= 2\rho +\frac{1}{2} \ln U_1 V_1 - \ln 8,
\label{psi}
\ee
the second order equation, not surprisingly, 
can in turn be rewritten in the Liouville form
\be
4\partial_{u} \partial_{v} \psi = e^{2\psi}
\ee
whose general solution is
\be 
\psi (u,v)= \frac{1}{2}\ln (4 \lambda^4 l'(u) l'(v)) - \ln (\frac{M}{\lambda}- \lambda^2 l(u) l(v))
\label{liou}
\ee
with $l(u)$ and  $l(v)$ the real restrictions of
an arbitrary analytic function $l(z)$ of the complex variable $z$ 
and with $M$ and $\lambda$ two real integration constants.
We can now combine Eqs.~(\ref{psi}) and (\ref{liou}) to write the general 
solution of Eq.~(\ref{mircea}) in the form
\be
\rho (u,v)= \frac{1}{2} \ln \Bigl[\frac{U(u)V(v)}{\frac{M}{\lambda}- \lambda^2 l(u) l(v)}\Bigr],
\ee
where 
\be
U(u)=\frac{4\lambda}{\sqrt{l'(u) U_1(u)}}, ~~~~~ V(v)=\frac{4\lambda}{\sqrt{l'(v) V_1(v)}}
\ee
are two new arbitrary functions replacing those introduced in 
Eq.~(\ref{orig}).

Changing coordinates to
\be
\hat{u}=l(u), ~~~~~~ \hat{v}=l(v),
\ee
and performing a suitable residual Weyl transformation of the type 
(\ref{weyl}), we end up with the familiar Witten metric (\ref{me}), (\ref{WW}).

Finally, for the metric (\ref{me}), the remaining field equations, 
$\ein[{(0)}]{u}{u}=0,~ \ein[{(0)}]{v}{v}=0$, lead to relations between the,
so far arbitrary, functions $U, ~V,~ l$, to wit
\ba
\ln l'(u)&=& \ln U(u) +A_u \int^u U(t) dt \nonumber\\
\ln l'(v)&=& \ln V(v) +A_v \int^v V(t) dt.
\ea
These relations are obviously obeyed for the Witten metric.
\bigskip

\sectn{Smoothing the $R \ln R$ Lagrangian}

The pure gravitational (no dilaton !) action of Eq.~(\ref{ln}) already has part of the cutoff feature
we wish to exploit to the extent that the logarithm becomes 
complex if $R$ is negative.  By adding a term involving 
$\ln (a - R)$, we can further limit the
scalar curvature to 
the range $0 \leq R \leq a$.  Thus in Eq.~(\ref{mp}) we were led to consider 
\begin{equation}
\Agrav[(\beta)] = \int d^{2}x \sqrt{-g} R [\ln R + 
	\beta \ln (a-R)]. 
\label{eq:Agrav}
\end{equation}
Without loss of generality we can set $a~=~8$ here (this can always be achieved by simultaneous constant shifts in $\rho$ and $a$). The field equations now
take the form
\begin{equation}
\ein{\mu}{\nu} \equiv \half \met{\mu}{\nu} \left(R- \frac{\beta R^2}{8-R}\right) - \cov{\mu}\cov{\nu} I(R)
	+ \met{\mu}{\nu} \nabla^{\lambda}\cov{\lambda} I(R) = 0,
\label{eq:RlnRmott}
\end{equation}
where
\begin{equation}
I(R) = \ln R + \beta \ln (8 - R) - \frac{8 \beta}{8 - R}.
\label{eq:g}
\end{equation}
Requiring that $\rho$ depends only on $w = u v$,
we obtain, again first for the trace of the field equations, 
\begin{equation}
2 \left[\frac{8 \beta}{8 - R} - \beta - 1 \right] \dl{u} \dl{v} \rho
	+ \dl{u} \dl{v} I(R) = 0,
\label{eq:EL}
\end{equation}
where $R$ is given in terms of $\rho$ by Eq.~(\ref{R}).
We are interested in black hole-type solutions in which $R$ goes
asymptotically to zero as $w \rightarrow -\infty$, so let us
consider the following $\frac{1}{w}$ expansion of $\rho$:
\begin{equation}
\rho = -\half \ln(-\mu w) + \frac{A}{w} + \frac{B}{w^2}+...~,
\label{eq:rhoasy}
\end{equation}
where $A$ and $B$ are constants.
We then obtain the scalar curvature
\begin{equation}
R = -8 \frac{\mu}{w} \left( A + \frac{4 B - 2 A^2}{w} 
	+ O(1/w^2) \right),
\end{equation}
and 
\begin{equation}
I(R) = \ln \left( \frac{-8\mu A}{w} \right) 
	+ \frac{4 B - 2 A^2}{A w} + \frac{\mu A \beta}{w}
	- \beta \left( 1 - \frac{\mu A}{w} \right) + O(1/w^2).
\end{equation}
Inserting this into Eq.~(\ref{eq:EL}), we find that
\begin{equation}
B = \left( 1 - \frac{\mu \beta}{2} \right) A^2.
\end{equation}
Comparing this result to a $\frac{1}{w}$ expansion of the $\rho$ function for the Witten black hole, 
Eq.~(\ref{WW}),
we identify our  integration constants as $\mu = \lambda^2$ and 
$A = \frac{M}{2 \lambda^3}$.  In the
limit $\beta \rightarrow 0$, we recover the usual 2-d black
hole result 
$B = A^2$.  So as $w \rightarrow - \infty$ we have solutions that
behave like the usual black hole.

As already mentioned in Section 2, 
it is essential that the action in Eq.~(\ref{eq:Agrav}) have a term of the form
$\sqrt{-g} R \ln R$ which cannot be expanded around $R = 0$.  If we consider the simplest nontrivial
action $\int d^{2} x \sqrt{-g} R^2$ which can be expanded around $R = 0$,
we obtain the Euler-Lagrange equation
\begin{equation}
\dl{u} \dl{v} \left[ e^{-2 \rho} \dl{u} \dl{v} \rho \right] = e^{-2 \rho} (\dl{u} \dl{v} \rho)^2. 
\end{equation}
Inserting into this equation the Ansatz of Eq.~(\ref{eq:rhoasy}), we find that $A = 0$, and the only
asymptotically flat solution is pure 2-d Minkowski space.  Thus, were we to consider a more ``symmetric'' action by replacing $R \ln R$ by $R \ln (a + R)$, this action would not admit black hole solutions.
 
Since Eq.~(\ref{eq:EL}) is nonlinear (in fact nonpolynomial) 
in $R$, 
we cannot find an exact 
solution for $\rho$, and so we must be content with numerical 
methods if we
wish to extend these solutions in $\frac{1}{w}$.
If $\beta < 0$, $I(R)$ is a monotonically
increasing function ranging over the reals for 
$0 \leq R \leq 8$, and so we can numerically integrate 
Eq.~(\ref{eq:EL}) to obtain $I(R)$ and then solve 
Eq.~(\ref{eq:g}) for 
$R$.  If $\beta$ is small and $\mu A < 8$, deviations of the
solution given in Eq.~(\ref{eq:rhoasy})
from the usual black hole are negligible for large $|w|$.  As an
example, consider say $w = -20$.  Then we can determine the boundary
conditions by setting $\rho$, $d\rho/dw$, $R$, and $dR/dw$ equal 
to their usual black hole values for $w = -20$.  The results of such
an integration for 
\be
M = \lambda = 1,~~~~ \beta = -.001 \label{eq:num}
\ee
are presented
\begin{figure}
\setlength{\unitlength}{0.1bp}
\begin{picture}(3600,2160)(0,0)
\put(2850,1806){\makebox(0,0)[r]{$\rho$}}
\put(2850,1946){\makebox(0,0)[r]{R}}
\put(2008,0){\makebox(0,0){$w$}}
\put(350,1180){%
\makebox(0,0)[b]{\shortstack{$\rho$}}%
}
\put(3730,1180){%
\makebox(0,0)[b]{\shortstack{$R$}}%
}
\put(3161,151){\makebox(0,0){0}}
\put(2521,151){\makebox(0,0){-5}}
\put(1880,151){\makebox(0,0){-10}}
\put(1240,151){\makebox(0,0){-15}}
\put(600,151){\makebox(0,0){-20}}
\put(540,2109){\makebox(0,0)[r]{12}}
\put(3550,2109){\makebox(0,0)[r]{12}}
\put(540,1844){\makebox(0,0)[r]{10}}
\put(3550,1844){\makebox(0,0)[r]{10}}
\put(3550,1578){\makebox(0,0)[r]{8}}
\put(540,1578){\makebox(0,0)[r]{8}}
\put(3550,1313){\makebox(0,0)[r]{6}}
\put(540,1313){\makebox(0,0)[r]{6}}
\put(3550,1047){\makebox(0,0)[r]{4}}
\put(540,1047){\makebox(0,0)[r]{4}}
\put(3550,782){\makebox(0,0)[r]{2}}
\put(540,782){\makebox(0,0)[r]{2}}
\put(3550,516){\makebox(0,0)[r]{0}}
\put(540,516){\makebox(0,0)[r]{-0}}
\put(3550,251){\makebox(0,0)[r]{-2}}
\put(540,251){\makebox(0,0)[r]{-2}}
\end{picture}
\caption{Numerical solution for $\rho(w)$ obtained from Eq.~(\ref{eq:EL}) and the corresponding scalar curvature $R(w)$ in units in which the parameter $a$ 
in Eq.~(\ref{eq:Agrav}) is set to $a=8$.}
\label{graph}
\end{figure}
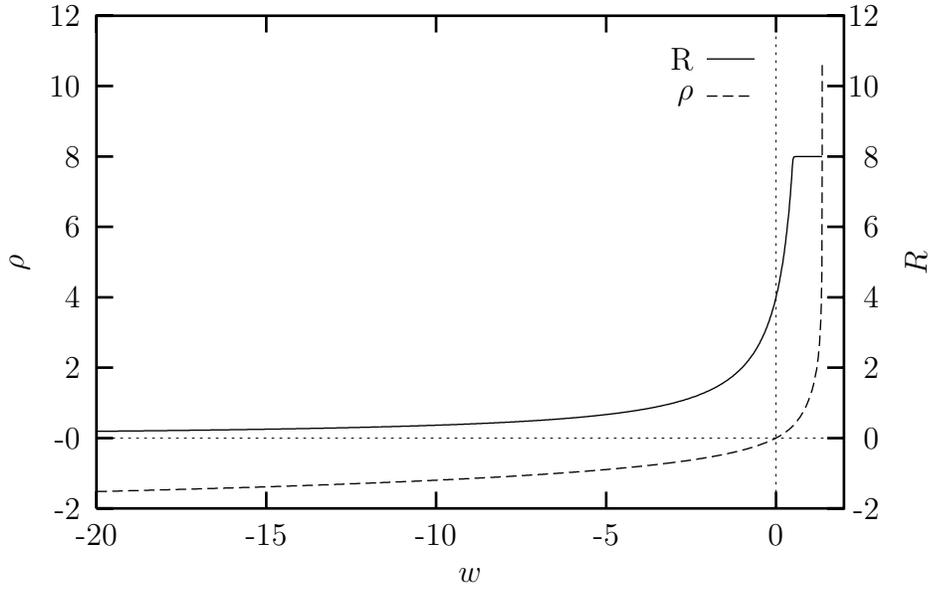
in Fig.~\ref{graph}.
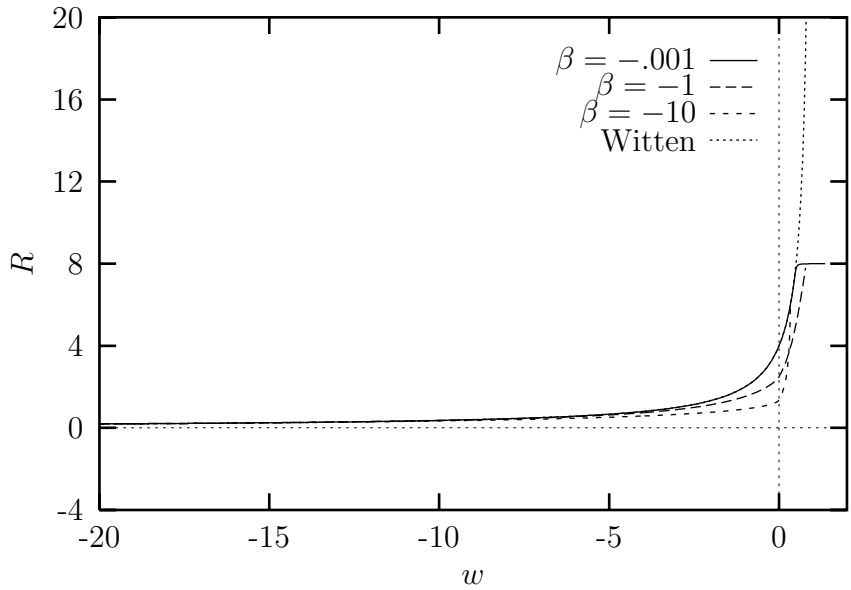
\begin{figure}
\setlength{\unitlength}{0.1bp}
\begin{picture}(3600,2160)(0,0)
\put(2850,1646){\makebox(0,0)[r]{Witten}}
\put(2850,1746){\makebox(0,0)[r]{$\beta=-10$}}
\put(2850,1846){\makebox(0,0)[r]{$\beta=-1$}}
\put(2850,1946){\makebox(0,0)[r]{$\beta=-.001$}}
\put(2008,0){\makebox(0,0){$w$}}
\put(350,1180){%
\makebox(0,0)[b]{\shortstack{$R$}}%
}
\put(3161,151){\makebox(0,0){0}}
\put(2521,151){\makebox(0,0){-5}}
\put(1880,151){\makebox(0,0){-10}}
\put(1240,151){\makebox(0,0){-15}}
\put(600,151){\makebox(0,0){-20}}
\put(540,2109){\makebox(0,0)[r]{20}}
\put(540,1799){\makebox(0,0)[r]{16}}
\put(540,1490){\makebox(0,0)[r]{12}}
\put(540,1180){\makebox(0,0)[r]{8}}
\put(540,870){\makebox(0,0)[r]{4}}
\put(540,561){\makebox(0,0)[r]{0}}
\put(540,251){\makebox(0,0)[r]{-4}}
\end{picture}
\caption{Numerical results for the curvature $R(w)$ at different values of $\beta$, compared to the Witten solution.}
\end{figure}

We can match this numerical result to approximate analytical 
solutions in certain convenient limits.  For small $w$, consider
the Ansatz
\begin{equation}
\rho = -\half \ln \gamma + L w + N w^2.
\label{eq:zero}
\end{equation}
Then we find
\begin{equation}
R(w) = 8 \gamma (L + (4 N - 2 L^2) w),
\label{eq:Rt0}
\end{equation}
and
\begin{equation}
N = L^2 \left[ 1+ \frac{\beta \gamma L}{2[(1-\gamma L)^2 - \beta \gamma L (2-\gamma L)]}\right] 
\end{equation}
In the limit of small $\beta$, this last result reduces to
\begin{equation}
N \approx L^2 \left[ 1 + 
	\frac{\gamma \beta L}{2 (1 - \gamma L)^2} \right].
\end{equation}
For the Witten black hole (\ref{WW}) the curvature at $w~=~0$ is $R(0)= 4\lambda^2$ (see Eq.~(2.4)), which for our numerical choice Eq.~(\ref{eq:num}) becomes $R(0)=4$. In Eq.~(\ref{eq:Rt0}) this then requires $\gamma L~\rightarrow~\frac{1}{2}$ for $\beta~\rightarrow~0$ so that to lowest order in $\beta$ this agrees with our numerical result. Here $w = 0$ still
corresponds to the event horizon of this smoothed black hole, and, 
for the above choice of parameters, the deviation of the curvature 
in Eq.~(\ref{eq:Rt0}) from the usual black hole curvature is 
proportional to
$\beta$ and so will be very small at the horizon.  Consequently,
an experimentalist living in this 2-d universe might have a 
difficult time indeed determining whether the laws of physics  admit singularities.  To do so he would actually have to 
pass through the event horizon, at which point he would be unable 
to publish his results. 

In Fig.~\ref{graph}, the function $\rho$ (not the curvature!)
appears to become singular near the point $w = 1.333$.  
Closer examination of the behavior of $\rho$ near
the point $w~=~\alpha$ at which $\rho$ is singular shows that 
\begin{equation}
\rho(w)  \approx - \ln \left( \frac{\alpha - w} {\sqrt{\alpha}} \right)
\label{eq:rhosing}
\end{equation}
there. The curvature $R$,
meanwhile, approaches its maximum allowed value $R~=~8$ along 
a linear trajectory as $w \rightarrow \alpha$.  If we include the next 
higher  order term
\begin{equation}
\rho(w) = - \ln \left( \frac{\alpha - w} {\sqrt{\alpha}} \right)
	+ D (\alpha - w) + O((\alpha - w)^2), \label{eq:rhojim}
\end{equation}
then we find this is indeed a solution to Eq.~(\ref{eq:EL}) with
$R = 8(1 - 2 D (\alpha - w))$.  

Like for the usual black hole, we have a coordinate singularity
at a finite value of $w$, but, unlike the usual black hole, the
scalar curvature remains finite at this point.  
It is readily seen that 
\begin{equation}
s = \frac{1}{\alpha - w}
\end{equation}
is an affine parameter along null geodesics for the metric
of Eq.~(\ref{eq:rhosing}).  Since $s \rightarrow \infty$ as
$w \rightarrow \alpha$, we conclude that the space-time is 
geodesically complete.  Instead of terminating in a curvature
singularity, as $s \rightarrow \infty$ the space-time is 
asymptotically de Sitter. This is similar to the situation described in  
Refs. \cite{B1, B2}. Notice that 
this is also quite similar to what happens in Born's electrodynamics, in which the electric field attains its maximum value at the
point that would correspond to the singularity in the Maxwell case.  
In the gravitational case as well, the scalar curvature attains its maximum 
value near the point which would have formerly corresponded to the 
curvature singularity.  
Only here, because of the absence of a space-time singularity, 
the space-time cannot end at a finite distance and the point of maximum 
curvature is pushed to an infinite distance from the 
event horizon.

The behavior of $\rho$ and $R$ for $\beta=-.001$ is representative
of what happens for small $\beta$ if we integrate Eq.~(\ref{eq:EL}) 
and use Eq.~(\ref{R})
to calculate the scalar curvature,
starting from the same boundary conditions.  In Fig. 1b we compare
graphs of the curvature for $\beta = -.001$, $\beta = -1$, $\beta = -10$, and 
for the Witten black hole ($\beta = 0$).

Finally, we have checked that the remaining field equations 
$\ein{u}{u}=0,~\ein{v}{v}=0$ are obeyed,as expected, 
near the points $w=0, w=\alpha$ and 
for large negative $w$, i.e. in the regions of validity of 
the expansions (\ref{eq:zero}), (\ref{eq:rhosing}), (\ref{eq:rhoasy}).

It should also be noticed that for {\em positive} $\beta$ the field equations
(\ref{eq:RlnRmott}) are solved by the de-Sitter metric with 
$R=\frac{8}{1+\beta}$. 

\sectn{ADM Masses for the $R \ln R$ Lagrangian}

Let us consider a space-time which for $w \rightarrow -\infty$
has the metric
\begin{equation}
ds^2 = - e^{2 \rho} du dv,
\end{equation}
where we assume $\rho$ takes the form of Eq.~(\ref{eq:rhoasy}).  
To calculate the ADM mass of this space-time, we need to work
in an asymptotically flat coordinate system, so let us
consider the asymptotic coordinate transformation
\begin{eqnarray}
u =&  -&\exp(\sqrt{\mu} (r - t)) \\
v =&  &\exp(\sqrt{\mu} (r + t)).
\end{eqnarray}
Then for large $r$,
\begin{equation}
ds^2 = (1 + 2 A e^{-2 \sqrt{\mu} r})(-dt^2 + dr^2),
\end{equation}
and 
\begin{equation}
R = 8 A \mu e^{- 2 \sqrt{\mu} r}.
\end{equation}

If we vary $\met{\mu}{\nu}$ in $\Agrav[(0)]$, we obtain the 
field equation
\begin{equation}
\ein[{(0)}]{\mu}{\nu} \equiv \half \met{\mu}{\nu} R 
	- \cov{\mu}\cov{\nu} \ln R
	+ \met{\mu}{\nu} \nabla^{\lambda}\cov{\lambda} \ln R = 0.
\label{eq:RlnRmot}
\end{equation}
As is well known, if the space-time possesses an asymptotic 
time-translation
symmetry, then we can write $\ein{0}{0} = \dl{1}m(r)$,
and the ADM mass of the space-time will be 
$m = \lim_{r \rightarrow \infty} m(r)$.

For a time-independent solution like the present metric, 
\begin{equation}
\ein[{(0)}]{0}{0} = \half \met{0}{0} R 
	+ \met{0}{0} \frac{1}{\sqrt{-g}} 
	\dl{1}(\sqrt{-g} g^{1 1} \dl{1} \ln R).
\end{equation}
The first term goes as $\exp(-2 \sqrt{\mu} r)$ and so does
not contribute to $m(r)$.  The remainder goes as
$-\dl{1}^2 (\ln R) + O(\exp(-2 \sqrt{\mu} r))$, and 
\begin{equation}
m(r) = -\dl{1}(-2 \sqrt{\mu} r + \ln(8 A)) = 2 \sqrt{\mu}.
\end{equation}
Thus the ADM mass for a black hole-type solution of 
$\Agrav[(0)]$
will be $2 \sqrt{\mu}$.  For the choice of constants
corresponding to the solution of Eq.~(\ref{WW}), 
$m = 2 \lambda$.  Keep in mind that $\lambda$ appears here as  a free parameter, i. e. an integration constant of this solution, and 
\textit{not as a cosmological constant} of the theory, as in the dilatonic gravity.  

Thus it is the constant term in Eq.~(\ref{eq:rhoasy}) which 
determines 
the ADM mass for 
the $R \ln R$ theory.  This is in contrast to  the original CGHS 
gravity, for which  the coefficient of $1/w$  determines the
mass.

If we consider $\Agrav[(\beta)]$ for $\beta \neq 0$, the ADM
mass due to the ``free'' action $\Agrav[(0)]$ will remain 
unchanged, but there may be an additional contribution
due to the perturbative term proportional to $\beta$.  The
gravitational field equation corresponding to this
perturbation is
\begin{eqnarray}
\ein[{\mathrm{int}}]{\mu}{\nu} & = &
	\frac{1}{\beta} \left(\ein{\mu}{\nu} 
	- \ein[{(0)}]{\mu}{\nu} \right)
	\equiv
	\frac{-R^2 \met{\mu}{\nu}}{2 (a - R)}
	- \cov{\mu} \cov{\nu}\left[ \ln(a-R) 
	- \frac{a}{a-R} \right] \nonumber \\
	& + & \met{\mu}{\nu} \cov{\lambda} \nabla^{\lambda}
	\left[\ln(a-R) - \frac{a}{a-R} \right] = 0.
\end{eqnarray}
The first term of $\ein[{\mathrm{int}}]{0}{0}$, 
being proportional to $R^2$, goes as 
$\exp(-4 \sqrt{\mu} r)$ and so does not contribute.  The other terms
reduce for large $r$ to
\begin{equation}
\ein[{\mathrm{int}}]{0}{0} = 
	2 (\cov{\mu} \cov{\nu} 
	- \met{\mu}{\nu} \cov{\lambda} \nabla^{\lambda})
	\left( \frac{R}{a} \right) + O(\exp(-4 \sqrt{\mu} r)),
\end{equation}
which will go as $\exp(-2 \sqrt{\mu} r)$.  Thus
the perturbation does not affect the ADM mass. This was expected since, to 
lowest order in $R$, the perturbation is topological.

\bigskip

\sectn{$R$-nonlinear 2-d dilaton gravity}\label{sec:mir}
For a metric of the standard  form (\ref{me}), 
the $R$-nonlinear dilaton gravity  action (\ref{np}) can be written as 
\be 
{\mathcal{A}}_F = \int du dv \sqrt{-g} e^{-2\phi}\left[4\lambda^2-16e^{-2\rho} \partial _u\phi \partial _v\phi + F(8\rca)\right]. \label{eq:action}
\ee
Here, for simplicity, we have  introduced the notation $\rca = R/8$ and $F$ is
the function defined in Eq.(\ref{fni}). By construction, 
$\mathcal{A}_F$ reduces to the CGHS action in the $\beta \rightarrow 0$ limit.
We should point out that by substituting the explicit conformal gauge form 
\ref{me} of the metric into the action, we limit ourselves to the trace of the
full set of gravitational equations. The remaining equations can then be dealt
with as in Sections 3 and 4. 

For the particular case when 
both $\rho$ and $\phi$ only depend on $w = uv$, the field equations following 
from the above action are 
\bleteq\be
e^{2\rho}\left[\lambda^2 + \gamma +\left(\frac{2}{\beta}-\frac{\gamma}{2}\right)\sqrt{1+\beta\rca} - \left(\frac{2}{\beta}+\frac{\gamma}{2}\right)\sqrt{1-\beta\rca}\right] - 4\dot{\phi}-4w\ddot{\phi}+4w\dot{\phi}^2= 0 
\ee
and
\be 
2f(\dot{\rho}+w\ddot{\rho})+2(f-2)(\dot{\phi}+w\ddot{\phi})-(\dot{f}+w\ddot{f})+4w\dot{\phi}^2(1-f)+4w\dot{\phi}\dot{f}=0
\ee
\eleteq
where 
\be
f(\rca)=\frac{d}{dR} F(R)= \frac{1}{8}\frac{d}{d\rca} F(8\rca)
\ee
and dot indicates derivative with respect to $w$. 
The "non-trace" gravitational field equations (obtained from the action (\ref {eq:action})) both reduce to the form:
\be
4\dot{\phi}^2(1-f)-\ddot{f} +4\dot{\phi}\dot{f}+2f\ddot{\phi}-4f\dot{\phi}\dot{\rho}+2\dot{\rho}\dot{f}=0.
\ee

Unlike in the CGHS case, these nonlinear equations cannot be solved analytically. However, for our purpose, it suffices to find an asymptotic solution for which $R~\rightarrow~0$ as $w~\rightarrow~-\infty$. This asymptotic solution will allow us 
to calculate the ADM mass of the black hole 
as in \cite{W}. 
With the known $\beta \rightarrow 0$ behavior, we look for an asymptotic solution of the following form:
\ba 
\rho&=&-\frac{1}{2}\ln {\rho_0} -n \ln w + \frac{1}{w} P\left(\frac{1}{w}\right) \nonumber \\
\phi&=&-k \ln w + \frac{1}{w} Q\left(\frac{1}{w}\right) \label{eq:previ}
\ea
where $P$ and $Q$ admit a Taylor series expansion in the (small) variable $\frac{1}{w}$. Inserting (\ref{eq:previ}) into the field equations, 
we obtain the solution:
\ba
\rho &=& -\frac{1}{2}\ln {(-\lambda^2w)} + \frac{1}{w}\frac{M}{2\lambda^3} +O\left(\frac{1}{w^2}\right) \nonumber \\
\phi &=& -\frac{1}{2} \ln w + \frac{1}{w}\frac{M}{2\lambda^3} +O\left(\frac{1}{w^2}\right) . \label{eq:sol1}
\ea

Since for small curvatures our lagrangian reproduces to leading 
order the CGHS lagrangian, it is clear that a solution of the type Eq. 
(\ref{eq:sol1}) exists. The curvature cut-off dependent departures from the 
CGHS solution appear in the higher order terms not displayed here, since they 
do not affect the value of the mass.
Although both the functions $\rho$ and $\phi$ become infinite
as $w \rightarrow -\infty$, just like in the CGHS case, physically relevant 
quantities, such 
as the metric $e^{2\rho}$ and the curvature, remain finite. The integration constant $M$  equals the mass of the black hole.  
Because of the form of $\mathcal {A}_F$, we must have $\rca \in (-\frac{1}{\beta}, \frac{1}{\beta})$, 
so that $R~(= 8\rca)$ is necessarily finite.  

As in Section 4, we expect a point $w=\alpha$ at which
the curvature $R$ attains its maximum allowed value 
$R=\frac{8}{\beta}$. Numerical solutions with such a behavior are
being studied.

\bigskip
\bigskip

\sectn{Discussion}

In this paper we have constructed gravitational analogues of Born's nonlinear 
electrodynamics and applied them to the black hole problem. 
This involves lagrangians which are non-polynomial in the components of the 
Riemann tensor. By construction, curvature singularities are then eliminated. 
Once inside the event horizon, a 
geodesic approaches for large values of the affine parameter 
a region in which the scalar curvature is constant and equal to its maximum
possible value, while space-time remains geodesically complete. 
We have treated here mainly the 2-d problem,
though higher dimensional cases can be treated in a similar manner, as
discussed briefly in Section 2. In the course of our arguments we were led to
study the 2-d $R\ln R$ lagrangian and its residual Weyl invariance.

Unlike the abelian and nonabelian Born-Infeld cases \cite{P, L, TS}, there is 
no string theory argument leading directly to the lagrangians introduced in 
Section 2. The appearance of the Born-Infeld lagrangian can be understood as
describing, through its point charge solutions,
the way fundamental strings attach to branes \cite{CM, G}.
It remains 
an interesting open problem to find the string-theoretic, or more generally
M-theoretic, role of the 
nonsingular black holes implied by the gravitational analogues introduced here.

\bigskip
\bigskip
\renewcommand{\thesection}{}
\sectn{Acknowledgements}
\renewcommand{\thesection}{\Alph{section}.}
We wish to thank Emil Martinec for a useful discussion and for calling the 
papers \cite{B1, B2} to our attention. This work was supported in part by 
NSF Grant No. PHY-9123780-A3. 

\bigskip

\appendix
\renewcommand{\thesection}{}
\sectn{Appendix}
\renewcommand{\thesection}{\Alph{section}.}
In this Appendix we will find the most general 2-d gravitational action 
\be 
{\mathcal A} = \int d^2x \sqrt{-g} K(R)
\ee
such that the Witten black hole Eqs. (\ref{me}), (\ref{WW}) solves the corresponding field equation. For metrics of the type (\ref{me}), we have, as was shown in Section 2, $R~=~8e^{-2\rho}\partial_u\partial_v\rho$. In terms of the variable $w = uv$, the function $\rho$ of Eq.~(\ref{WW}) is  
\be
e^{-2\rho}=\frac{M}{\lambda}-\lambda^2 w
\ee
and the curvature is found (see Eq.~(\ref{R})) to be 
\be 
R = \frac{4M\lambda^2}{M-\lambda^3w}.
\ee
The field equation following from the action (A.1) is 
\be
2e^{2\rho}K(R)-8(\dot{\rho}+w\ddot{\rho})K'(R)+4(\dot{K}'(R)+w\ddot{K}'(R)) = 0, \label{eq:eqro}
\ee
where the dot stands for derivative with respect to $w$ and the prime stands for derivative with respect to $R$. 

In terms of the new function 
\be
T(R) = R\frac{\rm d}{{\rm d}R}\Bigl(\frac{K(R)}{R}\Bigr) \label{eq:H}
\ee
the third order equation (\ref {eq:eqro}) becomes a second order differential equation
\be
\frac{T''(R)}{T'(R)} = -\frac{(3R-8\lambda^2)}{R(R-4\lambda^2)}
\ee
which can be easily integrated:
\be
T(R) = C_1\Bigl(\ln (R-4\lambda^2)-\ln R +\frac{4\lambda^2}{R}\Bigr)+C_2.
\ee
Inserting this into Eq.~(\ref{eq:H}) we can solve for $K(R)$ with the result
\ba
K(R) &=& C_1\Bigl[-\frac{R}{2}\bigl(\ln^2R+\ln^2(4\lambda^2)\bigr) - 4\lambda^2 +R\int \frac{\ln(R-4\lambda^2)}{R}{\rm d}R\Bigr]\nonumber \\ &+& C_2 R\ln R +C_3 R. \label{eq:ares}
\ea
The last term in the above result is topological and thus classically irrelevant. The $R\ln R$ lagrangian discussed in Section 3 is the second term. The 
term in square brackets, beyond representing an unwieldy transcendental function in $R$, does not share the ``residual'' Weyl invariance of the $R\ln R$ lagrangian, which was discussed in Section 3.

\end{document}